\documentclass[sigconf,10pt,natbib=false]{acmart}
\usepackage{float}
\usepackage{xcolor}
\usepackage{graphicx}
\usepackage[acronyms,nonumberlist,nopostdot,nomain,nogroupskip,acronymlists={hidden}]{glossaries}
\usepackage{xspace}
\usepackage{pgfplots}
\pgfplotsset{compat=1.18}
\usepackage{caption}
\usepackage{subcaption}
\newglossary[algh]{hidden}{acrh}{acnh}{Hidden Acronyms}
\AtBeginDocument{%
  }
\usepackage{pgfplots}
\pgfplotsset{compat=1.18}
\RequirePackage[
  datamodel=acmdatamodel,
  style=acmnumeric, 
  maxnames=5,
  ]{biblatex}

\usepackage{soul}
\usepackage{listings}
\usepackage{xcolor}

\begin{document}


\newcommand\note[2]{\color{#1}\bf #2}
\newcommand\af[1]{{\note{blue}{angelo: #1}}}
\newcommand\al[1]{{\note{green}{andrea: #1}}}
\newcommand\sm[1]{{\note{orange}{stefano: #1}}}
\newcommand\mpo[1]{{\note{purple}{michele: #1}}}
\newcommand\pb[1]{{\note{red}{paolo: #1}}}
\newcommand\tm[1]{{\note{red}{tommaso: #1}}}

\newcommand*{\phyl}{PHY-low\xspace}
\newcommand*{\pci}{\gls{pci}\xspace}
\newcommand*{\aiml}{\gls{ai}/\gls{ml}\xspace}
\newcommand{\ric}{\gls{ric}\xspace}
\newcommand{\rics}{\glspl{ric}\xspace}
\newcommand{\nearrt}{Near-\gls{rt}\xspace}
\newcommand{\nonrt}{Non-\gls{rt}\xspace}
\newcommand{\eax}{\gls{eaxcid}\xspace}
\newcommand{\ran}{\gls{ran}\xspace}

\newcommand{\smf}{xDevSM\xspace}

\definecolor{codegray}{rgb}{0.25,0.25,0.25} 
\definecolor{codepurple}{rgb}{0.58,0,0.82}

\lstdefinestyle{mystile-logs}{
  commentstyle=\color{gray},
  numberstyle=\tiny,
  basicstyle=\ttfamily\scriptsize,
  rulecolor=\color{black},
  breakatwhitespace=true,         
  breaklines=true,                 
  captionpos=b,
  frame=tb,
  keepspaces=true,                 
  numbers=left,                    
  numbersep=5pt,                  
  showspaces=false,                
  showstringspaces=false,
  showtabs=false,                  
  tabsize=2,
  xleftmargin=10pt,
  belowskip=-10pt,
}

\lstdefinelanguage{logs}{
  alsoletter={-},
  keywords={listen,on,off,dst,periodic},
  sensitive=false,
  comment=[l]{\#},
  morecomment=[s]{/*}{*/},
  moredelim=[l][\color{orange}]{\&},
  moredelim=[l][\color{magenta}]{*},
  morestring=[b]',
  morestring=[b]",
}


\title[\smf: A Flexible Framework for E2 Service Models]{\smf: Streamlining xApp Development With a Flexible Framework for O-RAN E2 Service Models}


\author[A. Feraudo, S. Maxenti, A. Lacava, P. Bellavista, M. Polese, T. Melodia]{Angelo Feraudo$^\dagger$,
Stefano Maxenti$^*$,
Andrea Lacava$^{*\ddagger}$,\\
Paolo Bellavista$^\dagger$,
Michele Polese$^*$, Tommaso Melodia$^*$
}
\affiliation{%
  \institution{$^\dagger$Department of Computer Science and Engineering, University of Bologna, Italy}
    \institution{$^*$Institute for the Wireless Internet of Things, Northeastern University, Boston, MA, U.S.A.}
    \institution{$^\ddagger$Sapienza University of Rome, Italy}
  \city{}
  \state{}
  \country{}
  }

\renewcommand{\shortauthors}{Feraudo et al.}
\newacronym{3gpp}{3GPP}{3rd Generation Partnership Project}
\newacronym{4g}{4G}{4th generation}
\newacronym{5g}{5G}{5th generation}
\newacronym{6g}{6G}{6th generation}
\newacronym{5gc}{5GC}{5G Core}
\newacronym{adc}{ADC}{Analog to Digital Converter}
\newacronym{aerpaw}{AERPAW}{Aerial Experimentation and Research Platform for Advanced Wireless}
\newacronym{ai}{AI}{Artificial Intelligence}
\newacronym{aimd}{AIMD}{Additive Increase Multiplicative Decrease}
\newacronym{am}{AM}{Acknowledged Mode}
\newacronym{amc}{AMC}{Adaptive Modulation and Coding}
\newacronym{amf}{AMF}{Access and Mobility Management Function}
\newacronym{aops}{AOPS}{Adaptive Order Prediction Scheduling}
\newacronym{api}{API}{Application Programming Interface}
\newacronym{apn}{APN}{Access Point Name}
\newacronym{ap}{AP}{Application Protocol}
\newacronym{aqm}{AQM}{Active Queue Management}
\newacronym{ausf}{AUSF}{Authentication Server Function}
\newacronym{avc}{AVC}{Advanced Video Coding}
\newacronym{awgn}{AGWN}{Additive White Gaussian Noise}
\newacronym{balia}{BALIA}{Balanced Link Adaptation Algorithm}
\newacronym{bbu}{BBU}{Base Band Unit}
\newacronym{bdp}{BDP}{Bandwidth-Delay Product}
\newacronym{ber}{BER}{Bit Error Rate}
\newacronym{bf}{BF}{Beamforming}
\newacronym{bler}{BLER}{Block Error Rate}
\newacronym{brr}{BRR}{Bayesian Ridge Regressor}
\newacronym{bs}{BS}{Base Station}
\newacronym{bsr}{BSR}{Buffer Status Report}
\newacronym{bss}{BSS}{Business Support System}
\newacronym{ca}{CA}{Carrier Aggregation}
\newacronym{caas}{CaaS}{Connectivity-as-a-Service}
\newacronym{cb}{CB}{Code Block}
\newacronym{cc}{CC}{Congestion Control}
\newacronym{ccid}{CCID}{Congestion Control ID}
\newacronym{cco}{CC}{Carrier Component}
\newacronym{cd}{CD}{Continuous Delivery}
\newacronym{cdd}{CDD}{Cyclic Delay Diversity}
\newacronym{cdf}{CDF}{Cumulative Distribution Function}
\newacronym{cdn}{CDN}{Content Distribution Network}
\newacronym{cli}{CLI}{Command-line Interface}
\newacronym{cn}{CN}{Core Network}
\newacronym{codel}{CoDel}{Controlled Delay Management}
\newacronym{comac}{COMAC}{Converged Multi-Access and Core}
\newacronym{cord}{CORD}{Central Office Re-architected as a Datacenter}
\newacronym{cornet}{CORNET}{COgnitive Radio NETwork}
\newacronym{cosmos}{COSMOS}{Cloud Enhanced Open Software Defined Mobile Wireless Testbed for City-Scale Deployment}
\newacronym{cots}{COTS}{Commercial Off-the-Shelf}
\newacronym{cp}{CP}{Control Plane}
\newacronym{cyp}{CP}{Cyclic Prefix}
\newacronym{up}{UP}{User Plane}
\newacronym{cpu}{CPU}{Central Processing Unit}
\newacronym{cqi}{CQI}{Channel Quality Information}
\newacronym{cr}{CR}{Cognitive Radio}
\newacronym{cran}{CRAN}{Cloud \gls{ran}}
\newacronym{crs}{CRS}{Cell Reference Signal}
\newacronym{csi}{CSI}{Channel State Information}
\newacronym{csirs}{CSI-RS}{Channel State Information - Reference Signal}
\newacronym{cu}{CU}{Central Unit}
\newacronym{d2tcp}{D$^2$TCP}{Deadline-aware Data center TCP}
\newacronym{d3}{D$^3$}{Deadline-Driven Delivery}
\newacronym{dac}{DAC}{Digital to Analog Converter}
\newacronym{dag}{DAG}{Directed Acyclic Graph}
\newacronym{das}{DAS}{Distributed Antenna System}
\newacronym{dash}{DASH}{Dynamic Adaptive Streaming over HTTP}
\newacronym{dc}{DC}{Dual Connectivity}
\newacronym{dccp}{DCCP}{Datagram Congestion Control Protocol}
\newacronym{dce}{DCE}{Direct Code Execution}
\newacronym{dci}{DCI}{Downlink Control Information}
\newacronym{dctcp}{DCTCP}{Data Center TCP}
\newacronym{dl}{DL}{Downlink}
\newacronym{dmr}{DMR}{Deadline Miss Ratio}
\newacronym{dmrs}{DMRS}{DeModulation Reference Signal}
\newacronym{drlcc}{DRL-CC}{Deep Reinforcement Learning Congestion Control}
\newacronym{drs}{DRS}{Discovery Reference Signal}
\newacronym{du}{DU}{Distributed Unit}
\newacronym{e2e}{E2E}{end-to-end}
\newacronym{earfcn}{EARFCN}{E-UTRA Absolute Radio Frequency Channel Number}
\newacronym{ecaas}{ECaaS}{Edge-Cloud-as-a-Service}
\newacronym{ecn}{ECN}{Explicit Congestion Notification}
\newacronym{edf}{EDF}{Earliest Deadline First}
\newacronym{embb}{eMBB}{Enhanced Mobile Broadband}
\newacronym{empower}{EMPOWER}{EMpowering transatlantic PlatfOrms for advanced WirEless Research}
\newacronym{enb}{eNB}{evolved Node Base}
\newacronym{endc}{EN-DC}{E-UTRAN-\gls{nr} \gls{dc}}
\newacronym{epc}{EPC}{Evolved Packet Core}
\newacronym{eps}{EPS}{Evolved Packet System}
\newacronym{es}{ES}{Edge Server}
\newacronym{etsi}{ETSI}{European Telecommunications Standards Institute}
\newacronym[firstplural=Estimated Times of Arrival (ETAs)]{eta}{ETA}{Estimated Time of Arrival}
\newacronym{eutran}{E-UTRAN}{Evolved Universal Terrestrial Access Network}
\newacronym{faas}{FaaS}{Function-as-a-Service}
\newacronym{fapi}{FAPI}{Functional Application Platform Interface}
\newacronym{fdd}{FDD}{Frequency Division Duplexing}
\newacronym{fdm}{FDM}{Frequency Division Multiplexing}
\newacronym{fdma}{FDMA}{Frequency Division Multiple Access}
\newacronym{fed4fire}{FED4FIRE+}{Federation 4 Future Internet Research and Experimentation Plus}
\newacronym{fir}{FIR}{Finite Impulse Response}
\newacronym{fit}{FIT}{Future \acrlong{iot}}
\newacronym{fpga}{FPGA}{Field Programmable Gate Array}
\newacronym{fr2}{FR2}{Frequency Range 2}
\newacronym{fs}{FS}{Fast Switching}
\newacronym{fscc}{FSCC}{Flow Sharing Congestion Control}
\newacronym{ftp}{FTP}{File Transfer Protocol}
\newacronym{fw}{FW}{Flow Window}
\newacronym{ge}{GE}{Gaussian Elimination}
\newacronym{gnb}{gNB}{Next Generation Node Base}
\newacronym{gop}{GOP}{Group of Pictures}
\newacronym{gpr}{GPR}{Gaussian Process Regressor}
\newacronym{gpu}{GPU}{Graphics Processing Unit}
\newacronym{gtp}{GTP}{GPRS Tunneling Protocol}
\newacronym{gtpc}{GTP-C}{GPRS Tunnelling Protocol Control Plane}
\newacronym{gtpu}{GTP-U}{GPRS Tunnelling Protocol User Plane}
\newacronym{gtpv2c}{GTPv2-C}{\gls{gtp} v2 - Control}
\newacronym{gw}{GW}{Gateway}
\newacronym{harq}{HARQ}{Hybrid Automatic Repeat reQuest}
\newacronym{hetnet}{HetNet}{Heterogeneous Network}
\newacronym{hh}{HH}{Hard Handover}
\newacronym{hol}{HOL}{Head-of-Line}
\newacronym{hqf}{HQF}{Highest-quality-first}
\newacronym{hss}{HSS}{Home Subscription Server}
\newacronym{http}{HTTP}{HyperText Transfer Protocol}
\newacronym{ia}{IA}{Initial Access}
\newacronym{iab}{IAB}{Integrated Access and Backhaul}
\newacronym{ic}{IC}{Incident Command}
\newacronym{ietf}{IETF}{Internet Engineering Task Force}
\newacronym{imsi}{IMSI}{International Mobile Subscriber Identity}
\newacronym{imt}{IMT}{International Mobile Telecommunication}
\newacronym{iot}{IoT}{Internet of Things}
\newacronym{ip}{IP}{Internet Protocol}
\newacronym{itu}{ITU}{International Telecommunication Union}
\newacronym{kpi}{KPI}{Key Performance Indicator}
\newacronym{kpm}{KPM}{Key Performance Measurement}
\newacronym{kvm}{KVM}{Kernel-based Virtual Machine}
\newacronym{los}{LOS}{Line-of-Sight}
\newacronym{lsm}{LSM}{Link-to-System Mapping}
\newacronym{lstm}{LSTM}{Long Short Term Memory}
\newacronym{lte}{LTE}{Long Term Evolution}
\newacronym{lxc}{LXC}{Linux Container}
\newacronym{m2m}{M2M}{Machine to Machine}
\newacronym{mac}{MAC}{Medium Access Control}
\newacronym{manet}{MANET}{Mobile Ad Hoc Network}
\newacronym{mano}{MANO}{Management and Orchestration}
\newacronym{mc}{MC}{Multi-Connectivity}
\newacronym{mcc}{MCC}{Mobile Cloud Computing}
\newacronym{mchem}{MCHEM}{Massive Channel Emulator}
\newacronym{mcs}{MCS}{Modulation and Coding Scheme}
\newacronym{mec}{MEC}{Multi-access Edge Computing}
\newacronym{mec2}{MEC}{Mobile Edge Cloud}
\newacronym{mfc}{MFC}{Mobile Fog Computing}
\newacronym{mgen}{MGEN}{Multi-Generator}
\newacronym{mi}{MI}{Mutual Information}
\newacronym{mib}{MIB}{Master Information Block}
\newacronym{miesm}{MIESM}{Mutual Information Based Effective SINR}
\newacronym{mimo}{MIMO}{Multiple Input, Multiple Output}
\newacronym{ml}{ML}{Machine Learning}
\newacronym{mlr}{MLR}{Maximum-local-rate}
\newacronym[plural=\gls{mme}s,firstplural=Mobility Management Entities (MMEs)]{mme}{MME}{Mobility Management Entity}
\newacronym{mmtc}{mMTC}{Massive Machine-Type Communications}
\newacronym{mmwave}{mmWave}{millimeter wave}
\newacronym{mpdccp}{MP-DCCP}{Multipath Datagram Congestion Control Protocol}
\newacronym{mptcp}{MPTCP}{Multipath TCP}
\newacronym{mr}{MR}{Maximum Rate}
\newacronym{mrdc}{MR-DC}{Multi \gls{rat} \gls{dc}}
\newacronym{mse}{MSE}{Mean Square Error}
\newacronym{mss}{MSS}{Maximum Segment Size}
\newacronym{mt}{MT}{Mobile Termination}
\newacronym{mtd}{MTD}{Machine-Type Device}
\newacronym{mtu}{MTU}{Maximum Transmission Unit}
\newacronym{mumimo}{MU-MIMO}{Multi-user \gls{mimo}}
\newacronym{mvno}{MVNO}{Mobile Virtual Network Operator}
\newacronym{nalu}{NALU}{Network Abstraction Layer Unit}
\newacronym{nas}{NAS}{Network Attached Storage}
\newacronym{nat}{NAT}{Network Address Translation}
\newacronym{nbiot}{NB-IoT}{Narrow Band IoT}
\newacronym{nfv}{NFV}{Network Function Virtualization}
\newacronym{nfvi}{NFVI}{Network Function Virtualization Infrastructure}
\newacronym{ni}{NI}{Network Interfaces}
\newacronym{nic}{NIC}{Network Interface Card}
\newacronym{nlos}{NLOS}{Non-Line-of-Sight}
\newacronym{now}{NOW}{Non Overlapping Window}
\newacronym{nsm}{NSM}{Network Service Mesh}
\newacronym[type=hidden]{nr}{NR}{New Radio}
\newacronym{nrf}{NRF}{Network Repository Function}
\newacronym{nsa}{NSA}{Non Stand Alone}
\newacronym{nse}{NSE}{Network Slicing Engine}
\newacronym{nssf}{NSSF}{Network Slice Selection Function}
\newacronym{o2i}{O2I}{Outdoor to Indoor}
\newacronym{oai}{OAI}{OpenAirInterface}
\newacronym{oaicn}{OAI-CN}{\gls{oai} \acrlong{cn}}
\newacronym{oairan}{OAI-RAN}{\acrlong{oai} \acrlong{ran}}
\newacronym{oam}{OAM}{Operations, Administration and Maintenance}
\newacronym{ofdm}{OFDM}{Orthogonal Frequency Division Multiplexing}
\newacronym{olia}{OLIA}{Opportunistic Linked Increase Algorithm}
\newacronym{omec}{OMEC}{Open Mobile Evolved Core}
\newacronym{onap}{ONAP}{Open Network Automation Platform}
\newacronym{onf}{ONF}{Open Networking Foundation}
\newacronym{onos}{ONOS}{Open Networking Operating System}
\newacronym{oom}{OOM}{\gls{onap} Operations Manager}
\newacronym{opnfv}{OPNFV}{Open Platform for \gls{nfv}}
\newacronym[type=hidden]{oran}{O-RAN}{Open \gls{ran}}
\newacronym{orbit}{ORBIT}{Open-Access Research Testbed for Next-Generation Wireless Networks}
\newacronym{os}{OS}{Operating System}
\newacronym{oss}{OSS}{Operations Support System}
\newacronym{pa}{PA}{Position-aware}
\newacronym{pase}{PASE}{Prioritization, Arbitration, and Self-adjusting Endpoints}
\newacronym{pawr}{PAWR}{Platforms for Advanced Wireless Research}
\newacronym{pbch}{PBCH}{Physical Broadcast Channel}
\newacronym{pcef}{PCEF}{Policy and Charging Enforcement Function}
\newacronym{pcfich}{PCFICH}{Physical Control Format Indicator Channel}
\newacronym{pcrf}{PCRF}{Policy and Charging Rules Function}
\newacronym{pdcch}{PDCCH}{Physical Downlink Control Channel}
\newacronym{pdcp}{PDCP}{Packet Data Convergence Protocol}
\newacronym{pdf}{PDF}{Probability Density Function}
\newacronym{pdsch}{PDSCH}{Physical Downlink Shared Channel}
\newacronym{pdu}{PDU}{Packet Data Unit}
\newacronym{pf}{PF}{Proportional Fair}
\newacronym{pgw}{PGW}{Packet Gateway}
\newacronym{phich}{PHICH}{Physical Hybrid ARQ Indicator Channel}
\newacronym{phy}{PHY}{Physical}
\newacronym{pmch}{PMCH}{Physical Multicast Channel}
\newacronym{pmi}{PMI}{Precoding Matrix Indicators}
\newacronym{powder}{POWDER}{Platform for Open Wireless Data-driven Experimental Research}
\newacronym{ppo}{PPO}{Proximal Policy Optimization}
\newacronym{ppp}{PPP}{Poisson Point Process}
\newacronym{prach}{PRACH}{Physical Random Access Channel}
\newacronym{prb}{PRB}{Physical Resource Block}
\newacronym{psnr}{PSNR}{Peak Signal to Noise Ratio}
\newacronym{pss}{PSS}{Primary Synchronization Signal}
\newacronym{pucch}{PUCCH}{Physical Uplink Control Channel}
\newacronym{pusch}{PUSCH}{Physical Uplink Shared Channel}
\newacronym{qam}{QAM}{Quadrature Amplitude Modulation}
\newacronym{qci}{QCI}{\gls{qos} Class Identifier}
\newacronym{qoe}{QoE}{Quality of Experience}
\newacronym{qos}{QoS}{Quality of Service}
\newacronym{quic}{QUIC}{Quick UDP Internet Connections}
\newacronym{rach}{RACH}{Random Access Channel}
\newacronym{ran}{RAN}{Radio Access Network}
\newacronym[firstplural=Radio Access Technologies (RATs)]{rat}{RAT}{Radio Access Technology}
\newacronym{rbg}{RBG}{Resource Block Group}
\newacronym{rcn}{RCN}{Research Coordination Network}
\newacronym{rc}{RC}{RAN Control}
\newacronym{rec}{REC}{Radio Edge Cloud}
\newacronym{red}{RED}{Random Early Detection}
\newacronym{renew}{RENEW}{Reconfigurable Eco-system for Next-generation End-to-end Wireless}
\newacronym{rf}{RF}{Radio Frequency}
\newacronym{rfc}{RFC}{Request for Comments}
\newacronym{rfr}{RFR}{Random Forest Regressor}
\newacronym{ric}{RIC}{RAN Intelligent Controller}
\newacronym{nrric}{Near-RT RIC}{Near-Real-Time RAN Intelligent Controller}
\newacronym{rlc}{RLC}{Radio Link Control}
\newacronym{rlf}{RLF}{Radio Link Failure}
\newacronym{rlnc}{RLNC}{Random Linear Network Coding}
\newacronym{rmr}{RMR}{RIC Message Router}
\newacronym{rmse}{RMSE}{Root Mean Squared Error}
\newacronym{rnis}{RNIS}{Radio Network Information Service}
\newacronym{rr}{RR}{Round Robin}
\newacronym{rrc}{RRC}{Radio Resource Control}
\newacronym{rrm}{RRM}{Radio Resource Management}
\newacronym{rru}{RRU}{Remote Radio Unit}
\newacronym{rs}{RS}{Remote Server}
\newacronym{rsrp}{RSRP}{Reference Signal Received Power}
\newacronym{rsrq}{RSRQ}{Reference Signal Received Quality}
\newacronym{rss}{RSS}{Received Signal Strength}
\newacronym{rssi}{RSSI}{Received Signal Strength Indicator}
\newacronym{rtt}{RTT}{Round Trip Time}
\newacronym{ru}{RU}{Radio Unit}
\newacronym{rw}{RW}{Receive Window}
\newacronym{rx}{RX}{Receiver}
\newacronym{s1ap}{S1AP}{S1 Application Protocol}
\newacronym{sa}{SA}{standalone}
\newacronym{sack}{SACK}{Selective Acknowledgment}
\newacronym{sap}{SAP}{Service Access Point}
\newacronym{sc2}{SC2}{Spectrum Collaboration Challenge}
\newacronym{scef}{SCEF}{Service Capability Exposure Function}
\newacronym{sch}{SCH}{Secondary Cell Handover}
\newacronym{scoot}{SCOOT}{Split Cycle Offset Optimization Technique}
\newacronym{sctp}{SCTP}{Stream Control Transmission Protocol}
\newacronym{sdap}{SDAP}{Service Data Adaptation Protocol}
\newacronym{sdk}{SDK}{Software Development Kit}
\newacronym{sdm}{SDM}{Space Division Multiplexing}
\newacronym{sdma}{SDMA}{Spatial Division Multiple Access}
\newacronym{sdl}{SDL}{Shared Data Layer}
\newacronym{sdn}{SDN}{Software-defined Networking}
\newacronym{sdr}{SDR}{Software-defined Radio}
\newacronym{seba}{SEBA}{SDN-Enabled Broadband Access}
\newacronym{sgsn}{SGSN}{Serving GPRS Support Node}
\newacronym{sgw}{SGW}{Service Gateway}
\newacronym{si}{SI}{Study Item}
\newacronym{sib}{SIB}{Secondary Information Block}
\newacronym{sinr}{SINR}{Signal to Interference plus Noise Ratio}
\newacronym{sip}{SIP}{Session Initiation Protocol}
\newacronym{siso}{SISO}{Single Input, Single Output}
\newacronym{sla}{SLA}{Service Level Agreement}
\newacronym{sm}{SM}{Service Model}
\newacronym{e2sm}{E2SM}{E2 Service Model}
\newacronym{e2ap}{E2AP}{E2 Application Protocol}
\newacronym{smf}{SMF}{Session Management Function}
\newacronym{smo}{SMO}{Service Management and Orchestration}
\newacronym{sms}{SMS}{Short Message Service}
\newacronym{smsgmsc}{SMS-GMSC}{\gls{sms}-Gateway}
\newacronym{snr}{SNR}{Signal-to-Noise-Ratio}
\newacronym{son}{SON}{Self-Organizing Network}
\newacronym{sptcp}{SPTCP}{Single Path TCP}
\newacronym{srb}{SRB}{Service Radio Bearer}
\newacronym{srn}{SRN}{Standard Radio Node}
\newacronym{srs}{SRS}{Sounding Reference Signal}
\newacronym{ss}{SS}{Synchronization Signal}
\newacronym{sss}{SSS}{Secondary Synchronization Signal}
\newacronym{st}{ST}{Spanning Tree}
\newacronym{svc}{SVC}{Scalable Video Coding}
\newacronym{tb}{TB}{Transport Block}
\newacronym{tcp}{TCP}{Transmission Control Protocol}
\newacronym{tdd}{TDD}{Time Division Duplexing}
\newacronym{tdm}{TDM}{Time Division Multiplexing}
\newacronym{tdma}{TDMA}{Time Division Multiple Access}
\newacronym{tfl}{TfL}{Transport for London}
\newacronym{tfrc}{TFRC}{TCP-Friendly Rate Control}
\newacronym{tft}{TFT}{Traffic Flow Template}
\newacronym{tgen}{TGEN}{Traffic Generator}
\newacronym{tip}{TIP}{Telecom Infra Project}
\newacronym{tm}{TM}{Transparent Mode}
\newacronym{to}{TO}{Telco Operator}
\newacronym{tr}{TR}{Technical Report}
\newacronym{trp}{TRP}{Transmitter Receiver Pair}
\newacronym{ts}{TS}{Technical Specification}
\newacronym{tti}{TTI}{Transmission Time Interval}
\newacronym{ttt}{TTT}{Time-to-Trigger}
\newacronym{tx}{TX}{Transmitter}
\newacronym{uas}{UAS}{Unmanned Aerial System}
\newacronym{uav}{UAV}{Unmanned Aerial Vehicle}
\newacronym{udm}{UDM}{Unified Data Management}
\newacronym{udp}{UDP}{User Datagram Protocol}
\newacronym{udr}{UDR}{Unified Data Repository}
\newacronym{ue}{UE}{User Equipment}
\newacronym{uhd}{UHD}{\gls{usrp} Hardware Driver}
\newacronym{ul}{UL}{Uplink}
\newacronym{um}{UM}{Unacknowledged Mode}
\newacronym{uml}{UML}{Unified Modeling Language}
\newacronym{upa}{UPA}{Uniform Planar Array}
\newacronym{upf}{UPF}{User Plane Function}
\newacronym{urllc}{URLLC}{Ultra Reliable and Low Latency Communications}
\newacronym{usa}{U.S.}{United States}
\newacronym{usim}{USIM}{Universal Subscriber Identity Module}
\newacronym{usrp}{USRP}{Universal Software Radio Peripheral}
\newacronym{utc}{UTC}{Urban Traffic Control}
\newacronym{vim}{VIM}{Virtualization Infrastructure Manager}
\newacronym{vm}{VM}{Virtual Machine}
\newacronym{vnf}{VNF}{Virtual Network Function}
\newacronym{volte}{VoLTE}{Voice over \gls{lte}}
\newacronym{voltha}{VOLTHA}{Virtual OLT HArdware Abstraction}
\newacronym{vr}{VR}{Virtual Reality}
\newacronym{vran}{vRAN}{Virtualized \gls{ran}}
\newacronym{vss}{VSS}{Video Streaming Server}
\newacronym{wbf}{WBF}{Wired Bias Function}
\newacronym{wf}{WF}{Waterfilling}
\newacronym{wg}{WG}{Working Group}
\newacronym{wlan}{WLAN}{Wireless Local Area Network}
\newacronym{osm}{OSM}{Open Source \gls{nfv} Management and Orchestration}
\newacronym{pnf}{PNF}{Physical Network Function}
\newacronym{drl}{DRL}{Deep Reinforcement Learning}
\newacronym{mtc}{MTC}{Machine-type Communications}
\newacronym{osc}{OSC}{O-RAN Software Community}
\newacronym{mns}{MnS}{Management Services}
\newacronym{ves}{VES}{\gls{vnf} Event Stream}
\newacronym{ei}{EI}{Enrichment Information}
\newacronym{fh}{FH}{Fronthaul}
\newacronym{fft}{FFT}{Fast Fourier Transform}
\newacronym{laa}{LAA}{Licensed-Assisted Access}
\newacronym{plfs}{PLFS}{Physical Layer Frequency Signals}
\newacronym{ptp}{PTP}{Precision Time Protocol}
\newacronym{cbrs}{CBRS}{Citizen Broadband Radio Service}
\newacronym{rnti}{RNTI}{Radio Network Temporary Identifier}
\newacronym{tbs}{TBS}{Transport Block Size}

\newacronym{onr}{ONR}{Office of Naval Research}
\newacronym{afosr}{AFOSR}{Air Force Office of Scientific Research}
\newacronym{afrl}{AFRL}{Air Force Research Laboratory}
\newacronym{arl}{ARL}{Army Research Laboratory}

\newacronym{ct}{CT}{Continuous Testing}
\newacronym{mno}{MNO}{Mobile Network Operator}
\newacronym{oci}{OCI}{Open Container Initiative}
\newacronym{macsec}{MACsec}{Media Access Control Security}
\newacronym{pt}{PT}{Plain Text}
\newacronym{cuda}{CUDA}{Compute Unified Device Architecture}
\newacronym{dsp}{DSP}{Digital Signal Processing}

\newacronym{cus}{CUS}{Control, User, Synchronization}
\newacronym{dpd}{DPD}{Digital Pre-Distorsion}
\newacronym{cfr}{CFR}{Crest Factor Reduction}
\newacronym{pci}{PCIe}{Peripheral Component Interconnect Express}
\newacronym{dpu}{DPU}{Data Processing Unit}
\newacronym{rfsoc}{RFSoC}{Radio Frequency System-on-Chip}
\newacronym{if}{IF}{Intermediate Frequency}
\newacronym{nyu}{NYU}{New York University}
\newacronym{gh}{GH}{Grace Hopper}
\newacronym{trl}{TRL}{Technology Readiness Level}
\newacronym{srfa}{SRFA}{Special Research Focus Area}
\newacronym{qsfp}{QSFP}{quad small form factor pluggable}
\newacronym{pse}{PSE}{Performance Specialized Engine}
\newacronym{cae}{CAE}{Cognitive Analysis Engine}
\newacronym{simd}{SIMD}{Single Instruction/Multiple Data}
\newacronym{rt}{RT}{Real-Time}
\newacronym{asm}{ASM}{Advanced Sleep Mode}
\newacronym{aoa}{AoA}{Angle of Arrival}
\newacronym{eaxcid}{eAxC\_ID}{extended Antenna-Carrier Identifier}
\newacronym{bwp}{BWP}{Bandwidth Part}
\newacronym{dfe}{DFE}{Digital Front-End}
\newacronym{spi}{SPI}{Serial Peripheral Interface}
\newacronym{gpio}{GPIO}{General Purpose Input/Output}
\newacronym{nco}{NCO}{Numerically Controlled Oscillator}
\newacronym{lo}{LO}{Local Oscillator}
\newacronym{lna}{LNA}{Low-Noise Amplifier}
\newacronym{pll}{PLL}{Phased-Locked Loop}
\newacronym{som}{SOM}{System-on-Module}
\newacronym{papr}{PAPR}{Peak-to-Average Power Ratio}
\newacronym{pcb}{PCB}{Printed Circuit Board}
\newacronym{gcpw}{GCPW}{Grounded Co-Planar Waveguide}
\newacronym{cnn}{CNN}{Convolutional Neural Network}
\newacronym{gmp}{GMP}{Generalized Memory Polynomial}
\newacronym{ngrg}{nGRG}{next Generation Research Group}
\newacronym{mrl}{MRL}{Manufacturing Readiness Level}
\newacronym{fr}{FR}{Frequency Range}

\newacronym{sbom}{SBOM}{Software Bill of Materials}
\newacronym{hbom}{HBOM}{Hardware Bill of Materials}
\newacronym{vex}{VEX}{Vulnerability Exploitability eXchange}
\newacronym{dos}{DoS}{Denial of Service}
\newacronym{sme}{SME}{Small-Medium Enterprise}

\newacronym{ulpi}{ULPI}{Uplink Performance Improvement}
\newacronym{oem}{OEM}{Original Equipment Manufacturer}
\newacronym{nsin}{NSIN}{National Security Innovation Network}
\newacronym{dod}{DoD}{Department of Defense}
\newacronym{arpu}{ARPU}{Average Revenue per User}
\newacronym{opex}{OPEX}{operational expenses}
\newacronym{txb}{TXB}{Transmit Beam}
\newacronym{cve}{CVE}{Common Vulnerabilities and Exposure}
\newacronym{arc}{ARC}{Aerial RAN CoLab}

\begin{abstract}

\glsunset{ran}

 \glspl{ric} are programmable platforms that enable data-driven closed-loop control in the O-RAN architecture. They collect telemetry and data from the \ran, process it in custom applications, and enforce control or new configurations on the \ran. Such custom applications in the \nearrt \ric are called xApps, and enable a variety of use cases related to radio resource management.  
Despite numerous open-source and commercial projects focused on the \nearrt \ric,
 developing and testing xApps that are interoperable across multiple \ran implementations is a time-consuming and technically challenging process. This is primarily caused by the complexity of the protocol of the E2 interface, which enables communication between the \ric and the \ran while providing a high degree of flexibility, with multiple \emph{\glspl{sm}} providing plug-and-play functionalities such as data reporting and \ran control. 
In this paper, we propose \smf, an open-source flexible framework for O-RAN service models, aimed at simplifying xApp development for the \gls{osc} \nearrt \gls{ric}. \smf reduces the complexity of the xApp development process, allowing developers to focus on the control logic of their xApps and moving the logic of the E2 service models behind simple \glspl{api}. 
We demonstrate the effectiveness of this framework by deploying and testing xApps across various \gls{ran} software platforms, including OpenAirInterface and srsRAN. This framework significantly facilitates the development and validation of solutions and algorithms on O-RAN networks, including the testing of data-driven solutions across multiple \ran implementations.
\end{abstract}




\maketitle

\glsresetall
\vspace{-.45cm}
\section{Introduction}%
\label{section:introduction}

The Open \gls{ran} vision, being defined into a network architecture by the O-RAN ALLIANCE, brings a fundamental paradigm shift in how networks are deployed and optimized, through closed-loop control of the \gls{ran} exercised by the so-called \glspl{ric}.
The current O-RAN specifications include two instances of the \gls{ric}, operating at different time scales and on different parameters. The \nonrt \gls{ric}, hosted in the network \gls{smo}, supports control loops with a time scale higher than 1\,s, and enforces policies or high-level configurations on the system. 
The \nearrt \gls{ric} performs radio resource management by directly tuning the configuration parameters of the \gls{ran} at a time scale between 10\,ms and 1\,s. 
These components host custom applications, called rApps and xApps, respectively, which implement the closed-loop control logic~\cite{polese2022understanding}.

In this paper, we focus on the \nearrt \ric. This component directly exercises radio resource management and thus holds the potential to significantly influence the performance of the \ran. Prior work has shown how dynamically configuring the \gls{ran} stack through closed-loop control exercised through xApps can lead to significant performance improvements in spectrum utilization, throughput, and user satisfaction~\cite{tsampazipandora,puligheddu2023semoran,zangooei2024flexible,irazabal2024tc}. Data-driven agents controlling and optimizing the radio resource allocation yield performance gains in metrics that are orthogonal (e.g., throughput and latency) and hence often challenging to jointly optimize~\cite{parvez2018survey}. The role of the \rics has been explored in the context of network slicing~\cite{tsampazipandora}, load balancing and handover~\cite{lacava2023programmable,coronado2022roadrunner}, traffic shaping~\cite{irazabal2024tc}, and spectrum sharing~\cite{smith2021ran}, among others~\cite{polese2022understanding}.

Such flexibility and capabilities have led to several open-source projects focused on the \nearrt \ric, as well as implementations of the \ran termination for the interface connecting the \nearrt \ric to the disaggregated \glspl{gnb}, i.e., the E2 interface~\cite{santos2024managing}. 
Among these, the \gls{osc} provides an open-source implementation of a \nearrt \ric~\cite{nearrtric-osc}, an xApp framework~\cite{ricapp-osc}, and an E2 termination~\cite{e2sim}. 
Popular open-source 5G \ran implementations also provide E2 terminations on their base station code base, e.g., for \gls{oai}~\cite{oai} and srsRAN~\cite{srsran2023oran}. 

While the availability of multiple projects on E2 and \gls{ric} has fostered experimental research on data-driven Open RAN solutions~\cite{bonati2023openran,johnson2021nexran,upadhyaya2023open}, it remains challenging to develop an xApp that works across components developed by different projects or vendors. The E2 interface itself is split into two protocols: an application protocol, or E2AP, which manages the connectivity between the RAN and the \ric, and the E2 \glspl{sm}, which builds functionalities on top of E2AP, e.g., to expose \glspl{kpm} or control the \ran parameters. While the E2AP implementations across the \ran projects discussed above interoperate with the \gls{osc} \nearrt \ric, the \glspl{sm} implementations often refer to different versions of the O-RAN specifications, with non-overlapping sets of features and capabilities. This leaves researchers and developers in the Open RAN space with limited options for developing and testing xApps. An additional challenge is represented by the complexity of the \gls{sm} implementation, with the serialization and deserialization of the data needing to abide by specifications for ASN.1 payloads.

To address this challenge, in this paper, we design, develop, and test \smf, a flexible and open-source framework to develop xApps for the \gls{osc} \nearrt \ric capable of interoperating with different E2SM implementations.\footnote{\url{https://github.com/wineslab/xDevSM}\\\url{https://openrangym.com/tutorials/xdevsm-tutorial}} \smf provides xApp developers with a \gls{sdk} with \glspl{api} that implement the steps defined in various E2SM protocols, part of the interaction between the xApp, the \gls{ric}, and, eventually, the E2 termination on the \ran. This removes the challenges associated with developing code to configure the E2SM messages and leaves the developer with the task of defining the application logic. \smf leverages the \gls{osc} \ric components and extends the xApp logic with a wrapper that can interface with different shared libraries implementing E2SM \gls{kpm}. We tested the framework with multiple open-source \gls{ran} implementations, including srsRAN, \gls{oai}, and NVIDIA \gls{arc}, validating the functionality as well as the flexibility of the proposed approach.

The remainder of the paper is organized as follows. Section~\ref{sec:xapp-bg} provides details and background on the O-RAN architecture, xApp frameworks and development, and summarizes the key challenges. Section~\ref{sec:smf} introduces \smf, with a discussion on its architecture and implementation. Section~\ref{sec:deployments} reviews the \ran implementations which have been successfully interoperated with \smf. Finally, we discuss lessons learned, conclude the paper, and suggest future work in Section~\ref{sec:discussion}.

 \vspace{-.2cm}
\section{xApp Development Background}
\label{sec:xapp-bg}

To ensure vendor interoperability across \gls{ran} and \ric implementations, the O-RAN ALLIANCE \gls{wg} 3 has developed specifications for multiple protocols that define E2 interface~\cite{oran-wg3-e2-gap}.
The first specification is the \gls{e2ap}~\cite{oran-wg3-e2-ap}, which is used for general management operations, such as the connection and disconnection of a \gls{du} or \gls{cu} with the \ric, as well as providing a list of RAN functions supported by the \ran nodes.
The \gls{e2sm} specifications~\cite{oran-wg3-e2-sm} define general elements for the \glspl{sm}, which express the semantics of the interactions between the xApps and the RAN nodes.
Each service model is featured in a document extending the general specification of the \gls{e2sm} with parameters related to the use case of interest.
Examples of these extensions are the E2SM KPM~\cite{oran-wg3-e2-sm-kpm}, for data collection, and the E2SM \gls{rc}~\cite{oran-wg3-e2-sm-rc}, which defines the protocol to perform \gls{rrm} actions such as traffic steering through handover management.

The O-RAN ALLIANCE, through the \gls{osc}, provides open-source implementations of the components of the O-RAN architecture, contributed by telecom and xApp vendors, researchers, and academia.
The \gls{osc} follows a release cycle that translates the O-RAN technical specifications into different versions of the OSC codebase, together with bug fixes and overall improvements of the infrastructure~\cite{Bimo2022}.

\begin{figure}
    \centering
    \includegraphics[width=0.60\linewidth]{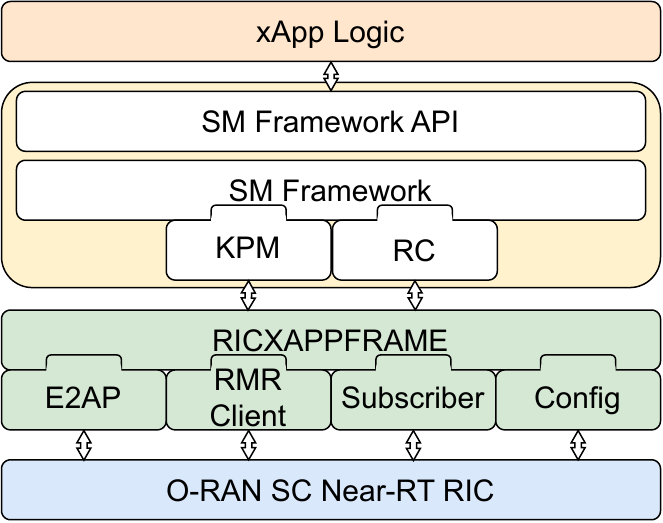}
    \caption{\smf and \ric components. The \smf components are in yellow, while the OSC \ric platform is in green and blue. The xApp developer only takes care of the xApp logic, in orange.}
    \Description[xApp Service Model Framework]{xApp Service Model Framework}
    \label{fig:xappsmframe}
    \vspace{-.6cm}
\end{figure}

In this context, the \gls{osc} has released a reference implementation for the \nearrt \ric, leveraging a micro-service architecture. This platform comprises various components, including internal message routing based on the \gls{rmr}, E2 termination, a subscription manager, a network information base database, and a shared data layer API among the others. Each of these components operates within a Kubernetes cluster, ensuring scalability and efficient resource management.
Besides the \nearrt \ric implementation, the \gls{osc} project also offers an xApp framework~\cite{ricapp-osc} that uses the \ric \gls{rmr} to manage the message exchange with the remaining components of the \ric platform.

Nevertheless, the rapid cycle with which the O-RAN ALLIANCE publishes new specifications often conflicts with the time required to develop new \gls{osc} releases, complicating the maintenance and update of the E2 specifications.
This has led to a flurry of scattered initiatives around E2 interface implementations. An example of this is represented by FlexApp paper~\cite{flexapp}, whose authors 
implement an E2* interface that is built on the E2 procedures but provides an additional layer of abstraction.
Other approaches have introduced simplified versions of the \glspl{sm} based on, for example, Protobuf buffers rather than ASN.1 data structures~\cite{moro-protobuf,villa2024x5gopenprogrammablemultivendor}. 
While this approach simplifies the development and testing of new \glspl{sm}, it is not fully O-RAN compliant and requires both xApp and \gls{ran} to support Protobuf. Additionally, encoding in Protobuf incurs higher overhead than using directly the binary field provided by the ASN.1 encoding~\cite{protobuf-comparison}.

The \ric platform itself has been subject to several development efforts, primarily for scaled-down versions of the Near-RT RIC, such as the srsRAN implementation~\cite{srsran2023oran} and the \gls{oai} FlexRIC. The srsRAN implementation, which is integrated with the srsRAN RAN components, provides the key functionalities of the original RIC at the I release version, but requires the deployment of the same \ric E2 libraries also on the \ran side, thus breaking the foundational concept of vendor interoperability. Compared to this approach, we provide a comprehensive framework that leverages a modular approach on the xApp side, facilitating scalability to different RAN implementations. The \gls{oai} FlexRIC is another SDK for software-defined-RAN controllers based on an E2-compatible protocol. This SDK includes an iApp component, which implements \glspl{e2sm} and exposes information to the xApp, and a server library, which multiplexes connections between xApps and dispatches \gls{e2ap} messages to E2 Nodes. The FlexRIC approach relies on a monolithic architecture, where all \nearrt \ric operations are included in a single component, decreasing the flexibility of the \ric platform. 
Moreover, \gls{oai} FlexRIC E2 Agent and \gls{osc} interoperability has not been tested properly, leading to incompatibility between \gls{oai} \ran and the \gls{osc} \nearrt \ric. 

This paper relies on the O-RAN components provided by the \gls{osc}, as these serve as a reference point that closely aligns with the architecture and specifications developed by the O-RAN ALLIANCE.

\begin{figure*}[t]
    \centering
    \includegraphics[width=.94\linewidth]{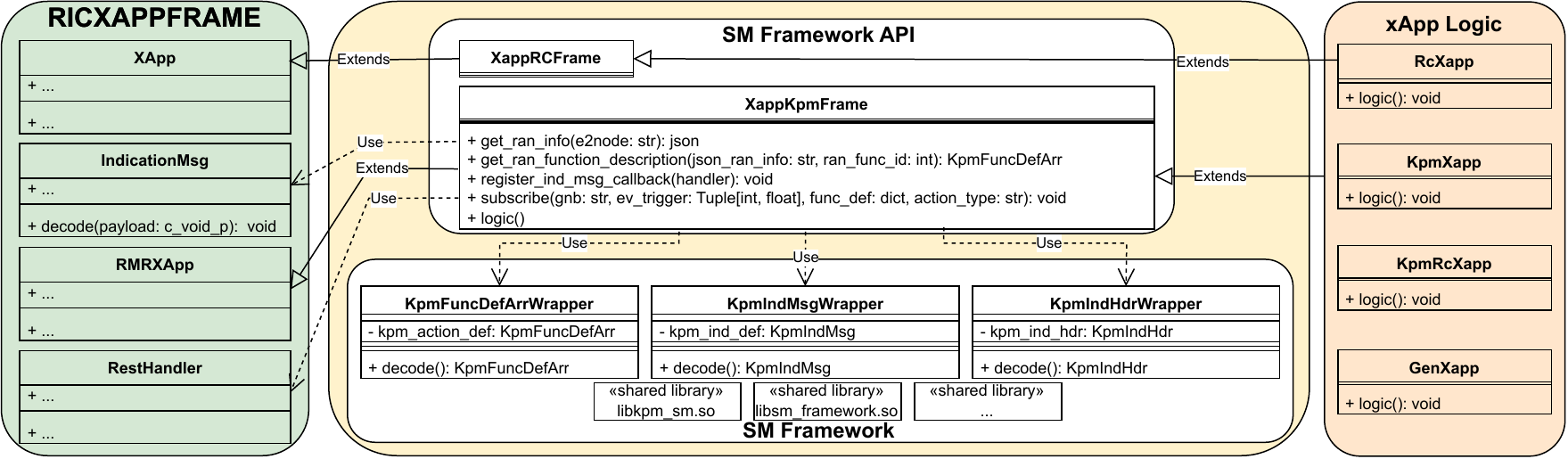}
        \vspace{-.3cm}
    \caption{Class diagram for \smf.}
    \Description[SM Framework Class Diagram]{SM Framework Class Diagram}
    \label{fig:xappframeclass}
    \vspace{-.2cm}
\end{figure*}

 \vspace{-.1cm}
\section{\smf: A Flexible \gls{sm} Framework for xApp Development}
\label{sec:smf}

In this section, we first review the architecture of \smf (Section~\ref{sec:archdes}), and then discuss its design and implementation in the context of the OSC \ric framework (Section~\ref{sec:implins}).
 \vspace{-.1cm}
\subsection{\smf Architecture}\label{sec:archdes}

As mentioned in previous sections, the \gls{osc} provides a comprehensive suite of libraries and functions aligned with their \nearrt platform version to support xApp development. This suite includes a message-level API for the \gls{rmr}, featuring callback registration, \gls{e2ap} encoding and decoding, subscription APIs, and the \gls{sdl} interface. In our scenario, we utilized the Python-based framework~\cite{oranscframepy}, which introduces two primary types of xApps: Reactive xApps (\texttt{RMRXApp}) and general xApps (\texttt{XApp}). 
Reactive xApps respond exclusively to incoming \gls{rmr} messages by invoking the appropriate callback based on the message type, with the main xApp capable of registering multiple callbacks. 
On the other hand, general xApps require the main xApp to actively fetch messages from the \gls{rmr} buffer.

The frameworks provided by \gls{osc} focus on the communication aspect within \ric components. However, once the xApp correctly subscribes or connects to the E2 node, there is a lack of easy-to-use functions to interact with the \gls{e2sm} (and the RAN functions) exposed by the E2 nodes. Hence, the burden of handling the serialization and deserialization procedures related to the \gls{sm} exposed by E2 nodes falls on the xApp developers. This requires them to know the specific \gls{e2sm} versions supported by the E2 node to implement a fully working xApp.

To address this challenge, we designed \smf, a \gls{sm} framework that enables encoding and decoding of \gls{e2sm} data in a plug-and-play fashion and through simple and well-defined \glspl{api}. As illustrated in Figure~\ref{fig:xappsmframe}, the framework is built on top of the Python xApp framework provided by the \gls{osc}~\cite{oranscframepy,oscxappframe} and comprises two components: the \gls{sm} Framework and the SM Framework API. 

The \gls{sm} Framework offers Python objects and functions that enable accurate decoding and encoding of messages related to supported \glspl{e2sm}. These functions are implemented in shared libraries written in C, which are dynamically loaded by the framework as needed. In this initial version, the shared libraries are generated using \gls{e2sm} functions defined in the Flex\ric project~\cite{flexric}, along with custom functions that simplify some xApp operations requiring data decoding. The Python objects within the framework handle the memory deallocation of the corresponding C structures by leveraging Python's garbage collector. To include new \glspl{sm}, developers can extend the shared libraries with the necessary encoding/decoding functions, and then add the corresponding Python objects.



The API component of the SM framework provides the necessary interfaces to interact with the \nearrt \ric and E2 Nodes and implement the xApp logic. It includes an abstract method that the xApp developer must define.
This method is called after the xApp setup and expresses the application-specific logic. For example, it can request the decoded \gls{ran} function description or store \glspl{kpi} in a time-series database. Additionally, the xApp containing the logic can receive decoded \gls{kpm} messages and can directly send \gls{rc} messages in an encoded form. These messages are managed using the \gls{sm} framework functionalities described above and sent to the \ric components using the \gls{osc} framework.

Compared to other solutions \cite{srsran2023oran}, \smf relies on shared libraries to manage ASN.1 serialization and deserialization procedures, leveraging the C language, which is the most widely adopted in this context~\cite{flexric,kpimon,bouncer}. This approach enhances the robustness of \smf, as it simplifies the process of finding and building shared libraries to support new service models. Additionally, the extensive use of C in this environment ensures better performance and compatibility, making \smf a more reliable and adaptable framework for xApp development.


\subsection{Design and Implementation of \smf}\label{sec:implins}
The new abstraction layer introduced in our framework facilitates data translation and optimizes memory management by handling the allocation and deallocation of external structures dynamically. As shown in Figure~\ref{fig:xappframeclass}, the API component provides a class named \texttt{XappKpmFrame}, which extends the \texttt{RMRXApp} class provided by the \gls{osc} xApp framework\cite{oscxappframe}. 
It does this by introducing methods to retrieve E2 node information, such as the node connection status to the \nearrt \ric and its available \ran functions, as well as methods related to the subscription procedure. Methods for \ran information retrieval use a Python object named \texttt{KpmFuncDefArrWrapper}. This wrapper references a custom data structure defined within the \texttt{libsm\_framework.so} shared library, which is essential for decoding and building the \textit{\ran Function Definitions}. Decoding these definitions is necessary as the \nearrt \ric subscription manager stores \ran function descriptors in hexadecimal and \gls{e2sm}-encoded formats. Furthermore, the wrapper handles memory deallocation related to the custom structure as soon as the Python object is no longer needed. 

The \texttt{subscribe} method of the \texttt{XappKpmFrame} class allows developers to specify the E2 node, the \ran functionalities to be monitored, and the reporting period for the xApp subscription. Before sending this information to the subscription manager, the \texttt{subscribe} method encodes the \ran functionalities and the reporting period using functions defined in the \texttt{libsm\_framework.so} shared library. Upon receiving this encoded information, the subscription manager performs an additional level of encoding using the \gls{e2ap} libraries and then forwards the subscription to the E2 nodes. It should be noted that the encoding functions provided by \texttt{libsm\_framework.so} utilize procedures defined in \texttt{libkpm\_sm.so}, generated using Flex\ric~\cite{flexric}, as these are closely related to the \gls{e2sm}. 

The xApp developers can use this framework by either extending the \texttt{XappKpmFrame} class or by encapsulating an instance of this class. Extending the \texttt{XappKpmFrame} offers more flexibility, allowing developers to customize behaviors of internal methods such as post-initialization actions, logger level adjustment, and modified decoding functions. On the other hand, encapsulating an instance of the \texttt{XappKpmFrame} class allows developers to utilize the class procedures without altering its internal behavior, simplifying integration and minimizing potential errors. 

In this paper, we focus on the first option as illustrated in Figure~\ref{fig:xappframeclass} (orange block). The \texttt{KpmXapp} class inherits from the \texttt{XappKpmFrame} and registers a callback method to manage \textit{Indication Messages} and defines the \texttt{logic} method. This ensures that all operations related to the dispatching of indication messages and decoding of the information are handled within the framework. Specifically, the framework automatically decodes the \gls{e2ap} part using the \texttt{RICXAPPFRAME IndicationMsg} class (green block Figure \ref{fig:xappframeclass}) and the \gls{e2sm} data using the \texttt{KpmIndMsgWrapper} and \texttt{KpmIndHdrWrapper} classes (yellow block Figure \ref{fig:xappframeclass}). The \gls{e2sm}-related classes expose a \texttt{decode} \gls{api}, which invokes a function defined in the \texttt{libkpm\_sm.so}. This function decodes the \gls{e2sm}-related information and returns a Python Object corresponding to the \gls{kpm} related data. This modular approach allows for easy updates to the \gls{e2sm} by simply changing the shared library version, provided that function signatures remain consistent across different \gls{e2sm} versions. Once the received message has been decoded, the registered function containing the behavior defined by the xApp developer is executed.

The \texttt{logic} method is an abstract method provided by the \texttt{XappKpmFrame} that allows the developers to define the xApp behavior. Within this method, developers can specify which E2 node to select based on a particular logic, identify \ran functionalities of interest, set reporting period, and more. We provide an example\footnote{\url{https://github.com/wineslab/xDevSM-xapps-examples}} of how an xApp can be defined using this framework, to demonstrate how it simplifies the process for developers when designing and implementing an xApp using existing frameworks, allowing them to focus exclusively on the application logic.

\section{Testing \smf with Open-Source \glspl{ran}}
\label{sec:deployments}

In this section, we showcase the flexibility and robustness of \smf by deploying an xApp based on our framework in combination with various real-world and simulated environments. We first consider \gls{oai}, in Section~\ref{subsec:oai}, and then we test srsRAN in Section~\ref{subsec:srs}. 

\subsection{OpenAirInterface}%
\label{subsec:oai}%
This section explores the effectiveness of our solution using software provided by \gls{oai} for the \ran, core, and \gls{ue}. We analyze a simulated scenario, using only \gls{oai}-based software, and a real-world scenario where the \gls{oai} \ran is paired with the Open5GS core and \gls{cots} \glspl{ue}.

\label{sec:oai-dep}
\begin{figure}[t]
    \centering
    \includegraphics[width=1\linewidth]{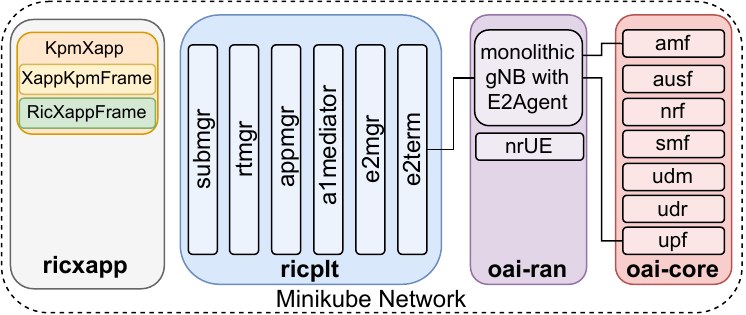}
    \caption{OpenAirInterface deployment in Minikube}
    \Description[OpenAirInterface deployment in a Minikube network]{OpenAirInterface deployment in Minikube network}
    \label{fig:oaib}
    \vspace{-.45cm}
\end{figure}

\begin{lstlisting}[float=b,language=logs,style=mystile-logs,
caption={Registration of the gNB on the \nearrt \ric},
label={lst:logs-gnb-ric}]
"inventoryName": "gnb_001_001_00000e05",
"globalNbId": {
"plmnId": "00F110",
"nbId": "00000000000000000000111000000101"
},
"connectionStatus": "CONNECTED"
\end{lstlisting}

\subsubsection{OpenAirInterface in a Box}
As illustrated in Figure~\ref{fig:oaib}, this deployment involves setting up a 5G network using Minikube, 
leveraging \gls{oai} for both \ran and core. To support the E2 interface and facilitate connectivity with the \nearrt \ric, the \gls{oai} \ran has been built with its default E2 agent, provided by FlexRIC project~\cite{flexric}. A pod in the \texttt{oai-ran} namespace implements an NR \gls{ue} and is connected to the base station using the OAI RFSim. This deployment creates an O-\ran-compliant environment in a single, isolated setup, enabling efficient testing and prototyping of xApps within a controlled environment. We utilize the \texttt{KpmXapp} described in the previous section to test its compatibility with OAI. In this configuration, the \gls{sm} framework utilizes the shared library (\texttt{libkpm\_sm.so}) produced during the building phase of the E2Agent from the FlexRIC SDK. We conducted several tests to confirm that \gls{oai} and the latest \gls{osc} \nearrt \ric Releases I and J can communicate effectively, while xApps perform their designated tasks. 
Listing~\ref{lst:logs-gnb-ric} shows a successful connection of a base station to the \nearrt \ric, whereas Listing~\ref{lst:logs-gnb} shows an example of the RAN function IDs and capabilities retrieved from the Base Station by the xApp and decoded by \smf. These capabilities include the amount of PDCP traffic exchanged in the measurement interval (\texttt{DRB.PdcpSduVolumeDL}, \texttt{DRB.PdcpSduVolumeUL}), the average throughput (\texttt{DRB.UEThpDl}, \texttt{DRB.UEThpUl}), and the number of \glspl{prb} allocated (\texttt{RRU.Prb\-TotDl}, \texttt{RRU.PrbTotUl}).

\begin{lstlisting}[float=t,language=logs,style=mystile-logs,
caption={Registration of the gNB on the xApp after decoding the ASN},
label={lst:logs-gnb}]
{[...], "id": "ricxappframe.xapp_frame", "mdc": {}, "msg": "Available functions: {0: [], 1: [], 2: [], 3: ['DRB.PdcpSduVolumeDL', 'DRB.PdcpSduVolumeUL', 'DRB.RlcSduDelayDl', 'DR
B.UEThpDl', 'DRB.UEThpUl', 'RRU.PrbTotDl', 'RRU.PrbTotUl'], 4: []}"}
{[...], "id": "ricxappframe.xapp_frame", "mdc": {}, "msg": "Selected functions: {3: ['DRB.PdcpSduVolumeDL', 'DRB.PdcpSduVolumeUL', 'DRB.UEThpDl', 'DRB.UEThpUl', 'RRU.PrbTotDl',
'RRU.PrbTotUl']}"}
{[...], "id": "ricxappframe.xapp_frame", "mdc": {}, "msg": "Preparing subscription for gnb: gnb_001_001_00000e05"}
{[...], "id": "ricxappframe.xapp_frame", "mdc": {}, "msg": "event trigger encoded: (8, 3, 231)"}
\end{lstlisting}

This deployment served as a base for developing, testing, and prototyping the framework proposed in this paper. All shared libraries created in this environment are reused in other deployments. 

\subsubsection{Over-the-Air OpenAirInterface Deployment with NVIDIA ARC}

\begin{figure}[t]
    \centering
    \includegraphics[width=\columnwidth]{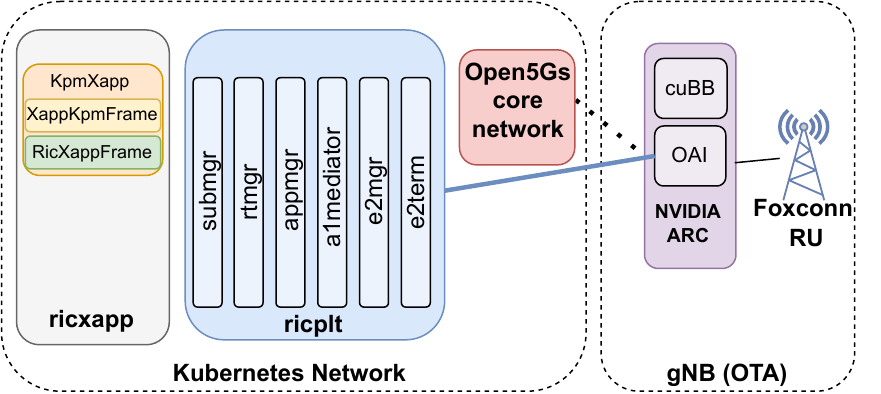}
    \caption{OpenAirInterface over the air}
    \Description[OpenAirInterface over the air]{OpenAirInterface over the air}
    \label{fig:oai-arc-ota}
    \vspace{-.6cm}
\end{figure}


Building on the simulated environment, we have extended our deployment to a real-world scenario that involves an over-the-air setup. In particular, we leverage the X5G testbed~\cite{villa2024x5gopenprogrammablemultivendor}, with \gls{oai} as the \gls{cu} and the \gls{du}-high, while the \gls{du}-low uses the NVIDIA \gls{arc} inline acceleration on \gls{gpu}. This is connected to a Foxconn RU operating in the 3.7-3.8 GHz band, with a bandwidth of 100 MHz, and connected to Samsung S23 and OnePlus AC2003 Nord \gls{cots} smartphones. In the same way as the previous setup, we test the default E2 agent for the OAI stack. Differently from the previous setup, though, we use Open5GS as a core network to validate functionalities with different \gls{amf} and \gls{upf} implementations, which may encode \gls{ue} information in slightly different ways. The final setup can be seen in Figure~\ref{fig:oai-arc-ota}.

We tested up to five connected \glspl{ue}, and the xApp correctly reports metrics for all of them. We also tested \glspl{ue} disconnections, which resulted in the correct removal from the list of available \glspl{ue} sent to the \nearrt \ric from the gNB. The \ran stack exposes the functions already shown in Listing~\ref{lst:logs-gnb}.

\begin{figure}
    \centering
    \begin{subfigure}[b]{0.49\columnwidth}
        \centering
        \begin{tikzpicture}
            \begin{axis}[
                xlabel={Timestamp [s]},
                ylabel={Throughput [Mbps]},
                grid=major,
                width=\columnwidth,
                legend style={at={(0.5,1.01)}, anchor=south},
                xtick={1,5,...,20}  
            ]
            \addplot coordinates {
                (1, 277)
                (2, 241)
                (3, 325)
                (4, 493)
                (5, 482)
                (6, 514)
                (7, 524)
                (8, 514)
                (9, 440)
                (10, 535)
                (11, 535)
                (12, 566)
                (13, 556)
                (14, 556)
                (15, 377)
                (16, 409)
                (17, 524)
                (18, 482)
                (19, 524)
                (20, 451)
            };
            \addlegendentry{iPerf3}
            
            \addplot coordinates {
                (1, 257.6444141)
                (2, 262.6076563)
                (3, 321.0705)
                (4, 501.7492031)
                (5, 489.9303438)
                (6, 528.5274375)
                (7, 531.4690547)
                (8, 524.7904766)
                (9, 449.6552422)
                (10, 551.0009688)
                (11, 530.8338672)
                (12, 576.9773047)
                (13, 560.9096406)
                (14, 568.7075313)
                (15, 396.8537813)
                (16, 412.3383125)
                (17, 525.6791406)
                (18, 498.5655781)
                (19, 533.0239063)
                (20, 463.6264609)
            };
            \end{axis}
        \end{tikzpicture}
        \caption{Downlink}
    \end{subfigure}
    \hfill
    \begin{subfigure}[b]{0.49\columnwidth}
        \centering
        \begin{tikzpicture}
            \begin{axis}[
                xlabel={Timestamp [s]},
                ylabel={Throughput [Mbps]},
                grid=major,
                width=\columnwidth,
                legend style={at={(0.5,1.01)}, anchor=south},
                xtick={1,5,...,20}  
            ]
            \addplot coordinates {
                (1, 70963.2/1024)
                (2, 103424/1024)
                (3, 103424/1024)
                (4, 103424/1024)
                (5, 109568/1024)
                (6, 102092.8/1024)
                (7, 100761.6/1024)
                (8, 110592/1024)
                (9, 112640/1024)
                (10, 115712/1024)
                (11, 117760/1024)
                (12, 116736/1024)
                (13, 113664/1024)
                (14, 97689.6/1024)
                (15, 91955.2/1024)
                (16, 103424/1024)
                (17, 108544/1024)
                (18, 106496/1024)
                (19, 108544/1024)
            };
            
            \addplot coordinates {
                (1, 66352.704/1024)
                (2, 106746.432/1024)
                (3, 103392.968/1024)
                (4, 104538.568/1024)
                (5, 111936.824/1024)
                (6, 105426.984/1024)
                (7, 102856.32/1024)
                (8, 112361.624/1024)
                (9, 114361.456/1024)
                (10, 116426.888/1024)
                (11, 122089.512/1024)
                (12, 116047.76/1024)
                (13, 118635.312/1024)
                (14, 102807.704/1024)
                (15, 93770.896/1024)
                (16, 103098.432/1024)
                (17, 111040.608/1024)
                (18, 108107.776/1024)
                (19, 111233.248/1024)
            };
            \legend{,xApp}
            \end{axis}
        \end{tikzpicture}
        \caption{Uplink}
    \end{subfigure}

    \caption{Comparison of iPerf3 (blue dots) and xApp (red squares) reported values with OAI}
\label{fig:oai-arc-thr}
\end{figure}
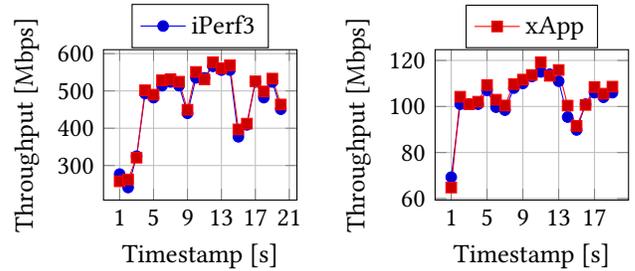

In Figure~\ref{fig:oai-arc-thr} we show that the values reported by the xApp are indeed tracking the traffic generated by the \gls{ue} using iperf3 over TCP, both in downlink and uplink. There is a slight offset of around 2\%, related to both a misalignment in reporting of the \glspl{kpm} by the xApp and iPerf3 logs, and the additional overhead introduced by headers in the TCP/IP stack and at PDCP, which is captured by the xApp and not by iPerf3. The latter indeed reports the application layer throughput, while the xApp measures throughput within the \ran. 

\subsection{srsRAN}%
\label{subsec:srs}%
This section analyzes the integration of \smf in a deployment based on srsRAN, to test the effectiveness of our framework with a different RAN software. Like the previous scenario, we first built a simulated environment using two virtual machines hosting the \nearrt \ric and the 5G components. The \ric virtual machine was also reused to test the \smf framework in a real-world scenario with over-the-air communications.

\subsubsection{srsRAN in a Box}
This deployment involves the simulated scenario provided in the srsRAN tutorial\footnote{\url{https://docs.srsran.com/projects/project/en/latest/tutorials/source/near-rt-ric/source/index.html}}, where the srsUE and the gNB communicate through a ZeroMQ-based RF driver and the deployed core network is Open5GS.

\begin{figure}[t]
    \centering
    \includegraphics[width=0.7\linewidth]{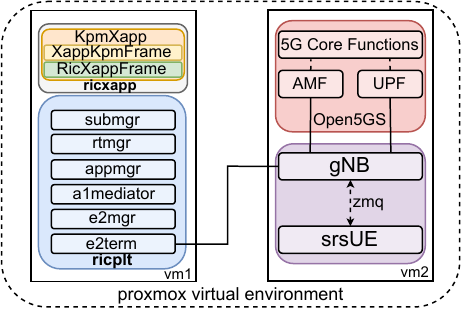}
    \caption{srsRAN simulated deployment}
    \Description[srsRAN simulated deployment]{srsRAN simulated deployment}
    \label{fig:srsranS}
    \vspace{-.5cm}
\end{figure}

As illustrated in Figure~\ref{fig:srsranS}, our setup differs from the one provided in the srsRAN tutorial in one way. We use two virtual machines within the same network: one that features all the 5G-related components, that is, 5G \ran, 5G \gls{ue}, and the 5G core, and the other hosts the \nearrt \ric Release J. The virtual machine running the \nearrt \ric is configured as a single-node Kubernetes cluster, where the E2Term SCTP service of the \nearrt \ric has been exposed to interact with the E2 node provided in srsRAN. For this test, we built the shared library for the service model \gls{kpm} v3.00 using the FlexRIC SDK, as srsRAN supports \gls{kpm} v3.00~\cite{oran-wg3-e2-sm-kpmv3}. We then deployed the xApp using our framework, exploiting the \gls{sm} designed for this environment. We notice that, by default, srsRAN exposes more \glspl{kpm} compared to OpenAirInterface.

\subsubsection{Over-the-Air srsRAN with 7.2 Split}

\begin{figure}[t]
    \centering
    \includegraphics[width=\columnwidth]{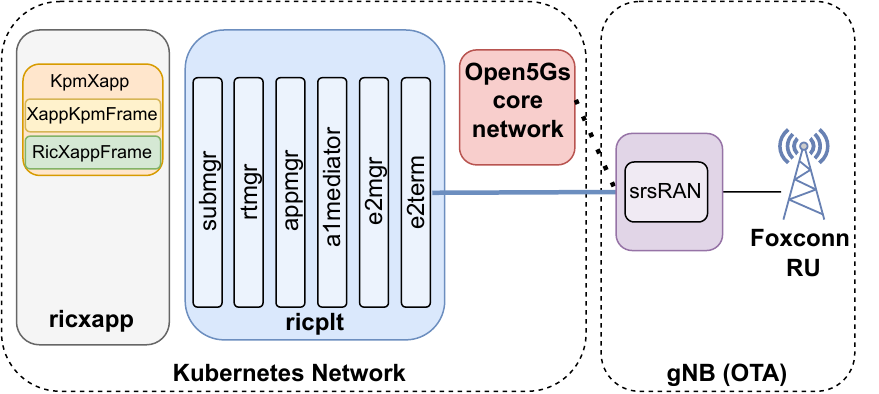}
    \caption{srsRAN over the air}
    \Description[srsRAN over the air]{srsRAN over the air}
    \vspace{-.3cm}
    \label{fig:srsran-ota}
\end{figure}

This deployment builds upon the setup from the previous scenario.
We use srsRAN with the same Foxconn RU as shown in Figure~\ref{fig:srsran-ota}. We transmit over a 40 MHz bandwidth and run the same iPerf3 tests as in the previous case. As expected, the xApp receives, decodes, and prints metrics from the gNB. 
%
Figure~\ref{fig:srsran-thr} compares the traffic generated by iPerf3 using TCP and the traffic calculated by the xApp. We notice a 1\% difference in downlink and around 10\% difference in uplink. This last difference is due to the fact that the throughput in uplink is limited to 10 Mbps and the xApp measures the performance at the PDCP level, therefore the overhead due the TCP/IP and PDCP headers is higher compared to other experiments. 
Finally, in some instances, the xApp fails to decode information associated to the update of the list of UEs, when UEs disconnect from the srsRAN base station. This does not happen with the \gls{oai} RAN implementation, thus further analysis on the srsRAN side is required. 





\section{Conclusions, Lessons Learned, and Future Extensions}
\label{sec:discussion}

In this paper, we proposed \smf, a flexible and open-source framework to develop xApps for the \gls{osc} \nearrt \ric. \smf introduces an abstraction layer to handle \gls{e2sm} encoding/decoding procedures, enabling xApp developers to focus on the control logic of their xApps and moving the service model logic behind simple \gls{api}. We proved the robustness and effectiveness of \smf by deploying xApps based on this framework in various environments with different \ran software. Our results demonstrate that \smf reduces the complexity of xApp development and facilitates the validation of solutions and algorithms in O-\ran networks.

The promising results shown in Section~\ref{sec:deployments} position the open-source framework as a platform with a significant potential for extensions and further development.  
At the time of writing, \smf only supports the \gls{e2sm} \gls{kpm} service model, with extensions planned for \gls{rc}. Additional extensions will be implemented for other service models as the O-RAN ALLIANCE WG3 defines them. 

To do this, and to change existing \glspl{e2sm}, \smf needs the design of new Python objects that match the corresponding newer and modified underlying structures (currently based on C). To address this for different \gls{e2sm} versions, the framework supports a set of custom structures that map the \gls{e2sm} notation structures to the Python objects supported by the framework. Nevertheless, this procedure is performed manually. The underlying C code used to match the \gls{sm} notation to the custom structure is manually modified, and then the shared library is rebuilt to reflect these changes. As part of our future work, we will consider automated approaches to achieve the same goal.

Additionally, the \gls{e2ap} encoding/decoding depends on the xApp framework provided by the \gls{osc}, and can get out of sync with respect to \ran E2AP implementations. Therefore, as part of our future work, we will track development in OSC and implement continuous testing solutions for \smf, the \nearrt \ric, and the \ran implementations considered in this paper.
\begin{figure}
    \centering
    \begin{subfigure}[b]{0.49\columnwidth}
        \centering
        \begin{tikzpicture}
            \begin{axis}[
                xlabel={Timestamp [s]},
                ylabel={Throughput [Mbps]},
                grid=major,
                width=\columnwidth,
                legend style={at={(0.5,1.01)}, anchor=south},
                xtick={1,5,...,20}  
            ]
            \addplot coordinates {
                (1, 189)
                (2, 210)
                (3, 199)
                (4, 231)
                (5, 241)
                (6, 210)
                (7, 220)
                (8, 220)
                (9, 220)
                (10, 231)
                (11, 231)
                (12, 220)
                (13, 252)
                (14, 231)
                (15, 241)
                (16, 252)
                (17, 210)
                (18, 231)
                (19, 273)
            };
            \addlegendentry{iPerf3}
            
            \addplot coordinates {
                (1, 186.0410156)
                (2, 207.4355469)
                (3, 210.3105469)
                (4, 228.9013672)
                (5, 236.9677734)
                (6, 219.2695313)
                (7, 219.1367188)
                (8, 223.7373047)
                (9, 234.2792969)
                (10, 228.9326172)
                (11, 236.1826172)
                (12, 234.0595703)
                (13, 242.7431641)
                (14, 235.8183594)
                (15, 253.6972656)
                (16, 254.3115234)
                (17, 210.6015625)
                (18, 239.0048828)
                (19, 265.2148438)
            };
            \end{axis}
        \end{tikzpicture}
        \caption{Downlink}
    \end{subfigure}
    \hfill
    \begin{subfigure}[b]{0.49\columnwidth}
        \centering
        \begin{tikzpicture}
            \begin{axis}[
                xlabel={Timestamp [s]},
                ylabel={Throughput [Mbps]},
                grid=major,
                width=\columnwidth,
                legend style={at={(0.5,1.01)}, anchor=south},
                xtick={1,5,...,20}  
            ]
            \addplot coordinates {
                (1, 9.5)
                (2, 10.8)
                (3, 10.7)
                (4, 9.34)
                (5, 10.3)
                (6, 10)
                (7, 9.54)
                (8, 10.8)
                (9, 9.74)
                (10, 9.63)
                (11, 10.3)
                (12, 9.58)
                (13, 10.1)
                (14, 9.29)
                (15, 10.7)
                (16, 10)
                (17, 10.1)
                (18, 10.5)
                (19, 9.85)
            };
            
            \addplot coordinates {
                (1, 10.62792969)
                (2, 12.5234375)
                (3, 11.68066406)
                (4, 10.90625)
                (5, 10.58398438)
                (6, 12.41113281)
                (7, 10.45507813)
                (8, 10.74707031)
                (9, 11.05664063)
                (10, 10.5625)
                (11, 10.74023438)
                (12, 9.091796875)
                (13, 12.62792969)
                (14, 11.57226563)
                (15, 10.53613281)
                (16, 13.12695313)
                (17, 11.95507813)
                (18, 11.28125)
                (19, 11.36328125)
            };
            \legend{,xApp}

            \end{axis}
        \end{tikzpicture}
        \caption{Uplink}
    \end{subfigure}

    \caption{Comparison of iPerf3 (blue dots) and xApp (red square) reported values with srsRAN}
    \vspace{-.3cm}
\label{fig:srsran-thr}
\end{figure}
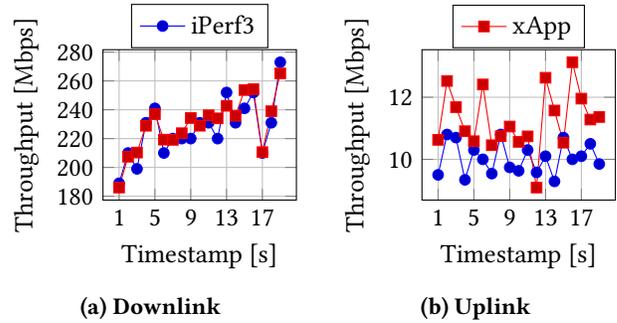
\begin{acks}
This article is based upon work partially supported by the O-RAN ALLIANCE, the U.S. National Science Foundation under grants CNS-2112471 and CNS-21200447, by the National Telecommunications and Information Administration (NTIA)'s Public Wireless Supply Chain Innovation Fund (PWSCIF) under Award No. 25-60-IF054, and by the National Recovery and Resilience Plan (Piano Nazionale di Ripresa e Resilienza, PNRR) RI SoBigData - Prot. IR0000013.
\end{acks}

\printbibliography

\end{document}